\input harvmac
%
%
\noblackbox
\def\npb#1#2#3{{\it Nucl.\ Phys.} {\bf B#1} (19#2) #3}
\def\plb#1#2#3{{\it Phys.\ Lett.} {\bf B#1} (19#2) #3}
\def\prl#1#2#3{{\it Phys.\ Rev.\ Lett.} {\bf #1} (19#2) #3}

\def\atmp#1#2#3{{\it Adv.\ Theor.\ Math.\ Phys.} {\bf #1} (19#2) #3}
\def\jhep#1#2#3{{\it JHEP\/} {\bf #1} (19#2) #3}
\newcount\figno
\figno=0
\def\fig#1#2#3{
\par\begingroup\parindent=0pt\leftskip=1cm\rightskip=1cm\parindent=0pt
\baselineskip=11pt
\global\advance\figno by 1
\midinsert
\epsfxsize=#3
\centerline{\epsfbox{#2}}
\vskip 12pt
{\bf Fig.\ \the\figno: } #1\par
\endinsert\endgroup\par
}
\def\figlabel#1{\xdef#1{\the\figno}}
\def\encadremath#1{\vbox{\hrule\hbox{\vrule\kern8pt\vbox{\kern8pt
\hbox{$\displaystyle #1$}\kern8pt}
\kern8pt\vrule}\hrule}}

\def\frac#1#2{{#1 \over #2}}

\def\p{\partial}
\def\semi{\subset\kern-1em\times\;}
\def\bar#1{\overline{#1}}

\def\p{\partial}
\def\pb{\bar{\partial}}  
\def\phib{\bar{\phi}}
\def\th{\theta}
\def\Cb{\bar{C}}
\def\ad{\bar a} 
\def\h{{1 \over 2}}
\def\Cb{\bar{C}}   
\def\Ab{\bar{A}}
\def\Db{\bar{D}}  
\def\phib{\bar{\phi}} 
\def\lt{\tilde{\lambda}}
\def\Tt{\tilde{\phi}}
\def\At{\tilde{A}}
\def\at{\tilde{a}}       
\def\IR{\relax{\rm I\kern-.18em R}}
%

\Title{\vbox{\baselineskip12pt
\hbox{hep-th/0010060}
\hbox{EFI-00-38}
\vskip-.5in}}
{\vbox{\centerline{Exact Noncommutative Solitons}}}
\medskip\bigskip
\centerline{Jeffrey A. Harvey, Per Kraus, and Finn Larsen}
\bigskip\medskip
\centerline{\it Enrico Fermi Institute and Department of Physics}
\centerline{\it  University of Chicago,} 
\centerline{\it  5640 S. Ellis Ave., 
Chicago, IL 60637, USA}
\medskip
\baselineskip18pt
\medskip\bigskip\medskip\bigskip\medskip
\baselineskip16pt
\noindent
We find exact solitons in a large class of  noncommutative
gauge theories using a simple solution generating technique.
The solitons in the effective field theory description of open string field
theory
are interpreted as D-branes for any value of the noncommutativity.
We discuss the vacuum structure of open string field theory in view of
our results.          

\Date{October, 2000}
\lref\david{
J.~R.~David,
``U(1) gauge invariance from open string field theory,''
hep-th/0005085.}
\lref\tatar{R. Tatar, ``A Note on Non-Commutative Field Theory and Stability of
Brane-Antibrane Systems,'' hep-th/0009213.}      
\lref\dmr{
K.~Dasgupta, S.~Mukhi and G.~Rajesh,
``Noncommutative tachyons,''
JHEP {\bf 0006}, 022 (2000)
[hep-th/0005006].}
\lref\schnabl{M. Schnabl, ``String field theory at large B-field and
noncommutative geometry,'' hep-th/0010034.}
\lref\ks{V .A. Kostelecky and S.Samuel, ``On a Nonperturbative Vacuum for the
Open Bosonic String,'' Nucl.Phys. {\bf B336} (1990) 263.}
\lref\sz{A. Sen and B. Zwiebach, ``Tachyon Condensation in String
Field Theory,'' hep-th/9912249.}
\lref\sendesc{A. Sen, ``Descent Relations Among Bosonic D-branes,'' Int.J. Mod.
Phys. {\bf A14} (1999) 4061, hep-th/9902105.}
\lref\sena{A.~Sen, ``Stable non-BPS bound states of BPS D-branes,''
\jhep{9808}{98}{010}, hep-th/9805019; 
``SO(32) spinors of type I and other solitons on brane-antibrane pair,''
\jhep{9809}{98}{023}, hep-th/9808141;
``Type I D-particle and its interactions,''
\jhep{9810}{98}{021}, hep-th/9809111; 
``Non-BPS states and branes in string theory,'' 
hep-th/9904207, and references therein.}
\lref\sennon{A. Sen, ``BPS D-branes on non-supersymmetric cycles,'' 
\jhep{9812}{98}{021}, hep-th/9812031.}
\lref\bergman{O.~Bergman and M.~R.~Gaberdiel,
``Stable non-BPS D-particles,'' \plb{441}{98}{133}, hep-th/9806155.}
\lref\hhk{J. A. Harvey, P.Horava and P. Kraus, ``D-Sphalerons and the Topology
of String Configuration Space,'' hep-th/0001143.}
\lref\berk{N. Berkovits, ``The Tachyon Potential in Open Neveu-Schwarz String
Field Theory,'' hep-th/0001084.}
\lref\polchinski{J.~Polchinski, ``Dirichlet-Branes and Ramond-Ramond 
Charges,'' \prl{75}{95}{4724}, hep-th/9510017.}
\lref\senspinors{A.~Sen, ``$SO(32)$ Spinors of Type I and Other Solitons on 
Brane-Antibrane Pair,'' \jhep{9809}{98}{023}, hep-th/9808141.}
\lref\phk{P. Ho\v rava, ``Type IIA D-Branes, K-Theory, and Matrix Theory,'' 
\atmp{2}{99}{1373}, hep-th/9812135.}
\lref\yi{P. Yi, ``Membranes from Five-Branes and Fundamental Strings from
D$p$-Branes,'' \npb{550}{99}{214}; hep-th/9901159.}
\lref\senpuz{A. Sen, ``Supersymmetric World-volume Action for Non-BPS
D-branes,'' \hfill\break hep-th/9909062.}
\lref\taylor{W. Taylor, ``D-brane effective field theory from string
field theory,'' hep-th/0001201.}
\lref\cft{C. G. Callan, I. R. Klebanov, A. W. Ludwig and J.M Maldacena,
``Exact solution of a boundary conformal field theory,'' Nucl.Phys.
{\bf B422} (1994) 417, hep-th/9402113;
J. Polchinski and L. Thorlacius, ``Free fermion representation of a boundary
conformal field theory,''Phys. Rev. {\bf D50} (1994) 622, hep-th/9404008;
P.Fendley, H. Saleur and N. P. Warner, ``Exact solution of a massless
scalar field with a relevant boundary interaction,'' Nucl. Phys.
{\bf B430} (1994) 577, hep-th/9406125;
A. Recknagel and V. Schomerus, ``Boundary deformation theory and
moduli spaces of D-branes,'' Nucl. Phys. {\bf B545} (1999) 233,hep-th/9811237.}
\lref\rs{L.Randall and R. Sundrum, ``An alternative to compactification,''
Phys.Rev.Lett. {\bf 83} (1999) 4690, hep-th/9906064.}
\lref\agms{M.~Aganagic, R.~Gopakumar, S.~Minwalla and A.~Strominger,
``Unstable solitons in noncommutative gauge theory'',
hep-th/0009142.}
\lref\dbak{D.~Bak,
``Exact multi-vortex solutions in noncommutative Abelian-Higgs theory'',
hep-th/0008204.}
\lref\jmw{D.~P.~Jatkar, G.~Mandal and S.~R.~Wadia,
``Nielsen-Olesen vortices in noncommutative Abelian Higgs model'',
JHEP {\bf 0009} (2000) 018 , hep-th/0007078.}
\lref\gms{R.~Gopakumar, S.~Minwalla and A.~Strominger,
``Noncommutative solitons'',
JHEP {\bf 0005} (2000) 020 , hep-th/0003160.}
\lref\ci{L.~Cornalba,
``D-brane physics and noncommutative Yang-Mills theory'',
hep-th/9909081; N.~Ishibashi,
``A relation between commutative and noncommutative descriptions of  
D-branes'', hep-th/9909176.}
\lref\witncsft{E.~Witten,
``Noncommutative tachyons and string field theory'',
hep-th/0006071.}
\lref\neksch{N.~Nekrasov and A.~Schwarz,
``Instantons on noncommutative $R^4$ and $(2,0)$ superconformal six  
dimensional theory'',
Commun.\ Math.\ Phys.\  {\bf 198} (1998) 689,
hep-th/9802068.}
\lref\harmor{J.~A.~Harvey and G.~Moore,
``Noncommutative tachyons and K-theory'',
hep-th/0009030.}
\lref\abs{M. F. Atiyah, R. Bott and A. Shapiro, ``Clifford Modules'',
Topology {\bf 3} suppl. 1 (1964) 3.}
\lref\absapp{E.~Witten,
``D-branes and K-theory'',
JHEP {\bf 9812} (1998) 019, hep-th/9810188;
P.~Horava,
``Type IIA D-branes, K-theory, and matrix theory'',
Adv.\ Theor.\ Math.\ Phys.\  {\bf 2} (1999) 1373,
hep-th/9812135.}
\lref\hklm{J.~A.~Harvey, P.~Kraus, F.~Larsen and E.~J.~Martinec,
``D-branes and strings as non-commutative solitons'',
JHEP {\bf 0007} (2000) 042,
hep-th/0005031.
}
\lref\ekawai{T. Eguchi and H. Kawai, ``Reduction of Dynamical Degrees
of Freedom in the Large N Gauge Theory,'' Phys. Rev. Lett. {\bf 48}
(1982) 1063.}
\lref\senunique{A. Sen, ``Uniqueness of Tachyonic Solitons,''
hep-th/0009090.}
\lref\somerefs{
A.~Sen and B.~Zwiebach,
``Large marginal deformations in string field theory,''
JHEP {\bf 0010}, 009 (2000)
[hep-th/0007153];
W.~Taylor,
``Mass generation from tachyon condensation for vector fields on  D-branes,''
JHEP {\bf 0008}, 038 (2000)
[hep-th/0008033];
A.~Iqbal and A.~Naqvi,
``On marginal deformations in superstring field theory,''
hep-th/0008127.
}
\lref\soch{C. Sochichiu, ``Noncommutative Tachyonic Solitons. Interaction
with Gauge Field,'' hep-th/0007217.}
\lref\wittenstrings{E. Witten, ``Overview of K Theory Applied to Strings,''
hep-th/0007175.}
\lref\nati{N. Seiberg, ``A Note on Background Independence in Noncommutative
Gauge Theories, Matrix Model and Tachyon Condensation,'' hep-th/0008013.}
\lref\pioline{B. Pioline and A. Schwarz, ``Morita equivalence and T-duality
(or B versus Theta),''
JHEP {\bf 9908} (1999) 021;hep-th/9908019.} 
\lref\truncate{A.~Sen and B.~Zwiebach,
``Tachyon condensation in string field theory'',
JHEP {\bf 0003} (2000) 002,
hep-th/9912249;
N.~Moeller and W.~Taylor,
``Level truncation and the tachyon in open bosonic string field theory'',
Nucl.\ Phys.\  {\bf B583} (2000) 105, hep-th/0002237.}
\lref\tsolrefs{J.~A.~Harvey and P.~Kraus,
``D-branes as unstable lumps in bosonic open string field theory'',
JHEP {\bf 0004} (2000) 012 ,
hep-th/0002117;
N.~Berkovits, A.~Sen and B.~Zwiebach,
``Tachyon condensation in superstring field theory'',
hep-th/0002211;
R.~de Mello Koch, A.~Jevicki, M.~Mihailescu and R.~Tatar,
``Lumps and p-branes in open string field theory'',
Phys.\ Lett.\  {\bf B482}, 249 (2000), hep-th/0003031;
N.~Moeller, A.~Sen and B.~Zwiebach,
``D-branes as tachyon lumps in string field theory'',
JHEP {\bf 0008}, 039 (2000),
hep-th/0005036.}
\lref\gmsii{R.~Gopakumar, S.~Minwalla and A.~Strominger,
``Symmetry restoration and tachyon condensation in open string theory'',
hep-th/0007226.}
\lref\senvac{A.~Sen,
``Some issues in non-commutative tachyon condensation'',
hep-th/0009038; A.~Sen,``Uniqueness of tachyonic solitons'',
hep-th/0009090.}
\lref\sw{
N.~Seiberg and E.~Witten,
``String theory and noncommutative geometry,''
JHEP {\bf 9909}, 032 (1999), hep-th/9908142.}
\lref\bsft{D.~Kutasov, M.~Marino and G.~Moore,
``Some exact results on tachyon condensation in string field theory'',
hep-th/0009148.}
\lref\witcub{E.~Witten,
``Noncommutative Geometry And String Field Theory'',
Nucl.\ Phys.\  {\bf B268} (1986) 253.}
\lref\witbsft{E.~Witten,
``On background independent open string field theory'',
Phys.\ Rev.\  {\bf D46} (1992) 5467, hep-th/9208027.}
\lref\gersh{A.~A.~Gerasimov and S.~L.~Shatashvili,
``On exact tachyon potential in open string field theory'',
hep-th/0009103.}
\lref\otherbsft{D.~Ghoshal and A.~Sen, ``Normalisation of the 
background independent open string field theory  action'', hep-th/0010021; 
L.~Cornalba,
``Tachyon Condensation in Large Magnetic Fields with Background 
Independent String Field Theory'', hep-th/0009191;
K.~Okuyama,
``Noncommutative Tachyon from Background Independent Open String Field 
Theory'', hep-th/0010028.}
\lref\cds{A.~Connes, M.~Douglas and A.~Schwarz, ``Noncommutative Geometry
and Matrix Theory:Compactification on Tori,'' JHEP {\bf 02} (1998) 003;
hep-th/9711162.}
\lref\schom{V.~Schomerus, ``D-Branes and Deformation Quantization'',
JHEP {\bf 9906} (1999) 030, hep-th/9903205.}
\lref\countb{U. Venugopalkrishna, ``Fredholm operators associated with strongly
pseudo convex domains in $C^n$,'' J. Functional Anal. {\bf 9}
(1972) 349;      
 L. Boutet de Monvel, ``On the index of
Toeplitz operators of several complex variables,'' Inv. Math.
{\bf 50} (1979) 249.}
\lref\dh{M. R. Douglas and C. Hull, ``D-branes and the Noncommutative
Torus,'' JHEP {\bf 9802} (1998) 008;hep-th/9711165.}
\lref\furuuchi{K.~Furuuchi,
``Equivalence of projections as gauge equivalence on noncommutative  space'',
hep-th/0005199; K. Furuuchi, ``Topological Charge of $U(1)$ Instantons
on Noncommutative $R^4$'' hep-th/0010006.}
\lref\pho{P.~Ho, ``Twisted bundle on noncommutative space and U(1) instanton'',
hep-th/0003012.}
\lref\grnek{D. J. Gross and N.A.Nekrasov, ``Monopoles and Strings
in Noncommutative Gauge Theory,'' JHEP {\bf 0007} (2000) 034; hep-th/
0005204; D. J. Gross and N. A. Nekrasov, ``Dynamics of Strings in 
Noncommutative Gauge Theory,'' hep-th/0007204.}
\lref\nekrasov{N. A. Nekrasov, ``Noncommutative Instantons Revisited,''
hep-th/0010017.}
\lref\poly{A.~P.~Polychronakos,
``Flux tube solutions in noncommutative gauge theories,''
hep-th/0007043.}
\lref\krs{P.~Kraus, A.~Rajaraman and S.~Shenker,
``Tachyon condensation in noncommutative gauge theory'',
hep-th/0010016.}
\lref\hkm{J.~A.~Harvey, D.~Kutasov and E.~J.~Martinec,
``On the relevance of tachyons,''
hep-th/0003101.}
\lref\senvanish{A.~Sen,
``Supersymmetric world-volume action for non-BPS D-branes,''
JHEP {\bf 9910}, 008 (1999)
[hep-th/9909062].}
\lref\dewit{B.~de Wit, J.~Hoppe and H.~Nicolai,
``On the quantum mechanics of supermembranes,''
Nucl.\ Phys.\  {\bf B305}, 545 (1988).}
\lref\coleman{See sec. 3.3 of S.Coleman, ``Quantum lumps and their classical
descendants,'' in {\it Aspects of Symmetry}, Cambridge University Press,
1985.}
\newsec{Introduction}

The study of semi-classical solutions of
noncommutative field theories has turned out to be an interesting
and rich subject with a variety of unexpected applications.
The properties of noncommutative $U(1)$ instantons provided
one of the early motivations for the study of noncommutative 
field theories \neksch, and it has been understood that
noncommutative gauge theory arises from a limit of string theory
\refs{\cds,\dh,\schom,\sw}.
More recently soliton solutions of scalar noncommutative  field theory
have been constructed \gms. These solutions have played an important 
role in constructing D-branes
as noncommutative solitons of the tachyon field of open string
theory \refs{\dmr,\hklm,\witncsft}.
There have also been many studies of solitons in noncommutative
scalar-gauge theory  \refs{\hklm,\grnek,
\poly,\jmw,\soch,\nati,\dbak,\agms,\tatar}.

In this paper we introduce a simple technique to find exact solutions 
in many different noncommutative gauge field theories, with or
without scalar fields.  We exploit it to
generate several new solutions, as well as some that were previously 
presented in the literature. Our method provides an efficient 
means of finding non-trivial solutions, and additionally
offers a unified interpretation of a large class of solitons.

The starting point of our solution generating technique is the
gauge invariance of noncommutative gauge theory. We then note 
that an enlarged class of transformations leave the field equations 
invariant by virtue of being ``almost gauge'',  without being 
full-fledged symmetries. These transformations thus map solutions to 
solutions, giving a simple method to generate new solutions. 
In this paper we describe the method for gauge field theories
and give various examples, including a detailed discussion of the  
effective field theory description of open string field theory.
We anticipate that generalizations of the construction can be
applied directly in the full string field theory, both in its
cubic formulation \witcub, and its background independent 
version \witbsft.

The solitons we construct are {\it exact} solutions for  all values
of the noncommutativity parameter, or equivalently the background
$B$-field. They are regular for any non-vanishing value of $B$, 
but singular in the limit $B\to 0$. They should therefore be 
distinguished from solutions which remain smooth as 
$B\to 0$; those form a seperate branch and are more difficult to 
construct explicitly. 
For example, some of the familiar $BPS$ solitons in commutative theories 
have generalizations to the noncommutative case \refs{\jmw}
but the solitons we
construct typically have higher energy than the corresponding
$BPS$ solitons; indeed we find that in some 
cases their energy diverges as $B\to 0$.

An important application of noncommutative solitons is to the 
description of tachyon condensation in open string field theory. 
Large noncommutativity $B=\infty$ introduces a large length scale
making derivatives in the action negligible; so in this limit it is 
simple to find $p$-brane solitons. Explicit computation shows that they 
have the same tension and other properties as $D$-branes; they should
therefore be identified with $D$-branes \hklm.
In this paper we generalize 
this result by exhibiting solitons with the same tension as $D$-branes 
for  all values of $B$. The new feature in the present construction
is that there is a non-vanishing gauge field excited in the solution, 
adjusted precisely so that the gauge covariant derivatives vanish
identically. This property effectively makes derivatives unimportant
without needing a new length scale to justify neglecting them.
Moreover, the form of the string field theory effective action 
is such that this feat can be accomplished at no cost in energy.

In our description D-branes are thus interpreted as solitons, 
with a width that 
depends on $B$ and vanishes as $B\to 0$. This result is satisfying
because $D$-branes are smaller than string scale in perturbative 
string theory at $B=0$. If the aim is to understand the solitons
at $B=0$ it may be useful
to interpret the noncommutativity as a regulator which can be
taken arbitrarily small. The physical picture is qualitatively similar 
to the one emerging from boundary string field theory 
\refs{\gersh,\bsft} (see also \otherbsft). There the 
boundary conformal field theory is perturbed by a relevant operator with the
mass parameter $u$. The soliton has width $~1/u$ and $u$ flows to 
$u=\infty$ in the exact description; thus there is a rough correspondance
$B\leftrightarrow 1/u$. In contrast, the description of D-brane solitons
in the level truncation approximation to the cubic string field theory
appears to yield solitons of finite width \tsolrefs. 
The field variables used 
in our discussion are evidently qualitatively similar to those of boundary 
string field theory, but not those of the level truncation scheme.

One of the goals of studying tachyon condensation in open string
field theory is to obtain a better understanding of the 
non-perturbative ``closed
string'' vacuum and its symmetries. In  open string theory
there is an obvious gauge symmetry changing $F$ and $B$
while fixing ${\cal F}=F+B$.  Thus the gauge invariant characterization 
of large noncommutativity is
${\cal F}=\infty$.  In the closed string vacuum far away from
any
D-branes there are no gauge fields to give rise to $F$, and different
constant values of $B$ are gauge equivalent. This 
gauge invariance is quite mysterious
from the open string point of view. Our results give a direct
calculational verification of the expected ${\cal F}$ independence
of $D$-brane properties. The relation of our results to other proposals
regarding the structure of the closed string vacuum will be
discussed in the final section of this paper.


The paper is organized as follows. 
In section 2 we present the solution generating technique for 
noncommutative solitons and give examples in simple noncommutative 
gauge theories.
In section 3 we use an effective field theory description of open string 
field theory to find solitons that we interpret as $D$-branes for
any value of ${\cal F}$. 
Section 4 discusses some implications of our results.

We note that several papers have exploited partial isometries
in the construction of instanton solutions 
\refs{\pho,\furuuchi,\nekrasov} 
and in a study of string field theory \schnabl.


\newsec{The Solution Generating Technique}
In many physical theories the equations of motion are left invariant
by a larger symmetry than is the full theory. This property is useful
for generating nontrivial solutions to the field equations: starting with a 
known (and typically simple) solution and acting with a symmetry of the 
equations of motion one finds a new (and often more involved) solution.
In the present section we apply this strategy to noncommutative gauge 
theories and consider as examples Yang-Mills-Higgs theories.
 The following section discusses the 
effective theories arising in open string field  theory.
In both cases, we will start with the vacuum and act with a solution generating
transformation to arrive at a soliton solution. 
The solutions we construct are exact for any value of noncommutativity.
One class of these solutions reduces to those of \gms\ in the limit of
large noncommutativity, while another class gives exact vortex solutions
and higher dimensional generalizations thereof.

\subsec{The General Construction}
A quantum field theory in a given dimension can always be represented
as a lower dimensional theory, with 
 the ``missing'' dimensions 
implemented as additional indices on the fields.  
This type of representation 
arises naturally in D-brane realizations of noncommutative field 
theories. Here D-branes can be represented as configurations of 
infinitely many lower dimensional D-branes. In this framework, the familiar 
$U(1)$ gauge theory on the brane becomes a  ``$U(\infty)$'' 
gauge symmetry in the lower dimensional theory and implements the 
invariance under area-preserving diffeomorphisms \refs{\dewit,\ci}.  

It is convenient to represent field configurations on the noncommutative space
 as operators acting on
an auxiliary Hilbert space ${\cal H}$  (see {\it e.g.} \gms.) 
In this formalism the symmetry is more precisely written as
a  $U({\cal H})$ symmetry of unitary transformations on  
Hilbert space transforming states and operators as
\eqn\gauget{\eqalign{
|\psi\rangle \to& U|\psi\rangle~,\cr
\langle \psi |\to& \langle \psi | \bar U~,\cr
{\cal O}\to&  U{\cal O} \bar U~.
}}
where $\bar{\cal O}$ denotes the Hermitian conjugate or adjoint of 
${\cal O}$.
In the operator representation the equations of motion can be arranged to
involve products of operators each transforming according to the
last equation in \gauget. They are clearly invariant under 
transformations satisfying
\eqn\runi{
\bar U U = I~.
}
That is, the condition \runi\ ensures that 
\eqn\runem{
{\delta S\over \delta {\cal O}}
\rightarrow U {\delta S\over \delta {\cal O}} \bar U~,}
and so will take solutions of the equations of motion to solutions.

True gauge transformations  of the  theory are realized by unitary 
operators satisfying $U \bar U=I$ as well as \runi. 
This should be distinguished from the present situation where the solution 
generating transformations are {\it not} full-fledged symmetries of the 
theory; in particular, they do not leave the action invariant. 
The condition \runi\ implies that $U \bar U$ is a projection
operator, so solution generating 
transformations
 are represented by the operators satisfying \runi\ but
\eqn\luni{
U\bar U = P~,
}
with $P$ a projection operator not equal to the identity operator.

The operator equation $O \bar O O = O$ implies that 
$O \bar O=P_1$ and $ \bar O O=P_2$ 
where $P_1,P_2$ are
projection operators.  Projection operators related in this way 
are said to be Murray-Von Neumann equivalent.  An operator obeying this 
equation is called a partial isometry. If $P_2$ is the identity 
operator then $O$ preserves inner products on the full Hilbert
space and is thus called 
an isometry. If $P_1$ is also the identity operator then $O$ is 
unitary.  Thus our generating transformations $U$ are non-unitary 
isometries and $P$ is Murray-von Neumann equivalent to the 
identity operator.

The solution generating transformations are  ``almost'' gauge
transformations.    For $P=I-P_n$, with $P_n$ a rank $n$ projection
operator, $U$ fails to be unitary only in an $n$ dimensional subspace.
In position space, the corresponding statement is that $U$ fails to be 
a true gauge transformation in a region of characteristic size 
$\sqrt{n \th}$ around the origin.  Thus acting with $U$ on the vacuum
will produce localized soliton solutions.  
Since $U$ does not generally commute with the Hamiltonian, the solitons
will have nonvanishing energy.
 These transformations
should not be confused with more familiar ``large'' gauge transformations.
The latter leave the classical action invariant but act nontrivially
on quantum states due to their nontrivial behavior at infinity.

It is often the case that the full Hilbert space factorizes into
several subspaces, {\it e.g.} corresponding to pairs of noncommutative
dimensions, or commutative ones. A more general solution generator 
can then be found which acts as a non-unitary isometry on 
each subspace {\it independently}. Apart from the obvious interest in 
generalizing the construction, this is useful because we are often
interested in solutions which do not depend on time, or other variables 
``along the brane''. Such solutions are generated by choosing the
identity transformation along the appropriate subspaces.  More generally,
we can take the transformations to act by taking states from one subspace
into those of another.  

It is not possible to realize \runi\ and \luni\ simultaneously 
in a finite dimensional vector space. To construct examples in an infinite 
dimensional separable Hilbert space, introduce an orthonormal 
basis $|k\rangle$, $k=0,1,\ldots$ and consider the shift operator
\eqn\shift{S = \sum_{k=0}^\infty |k+1\rangle \langle k|~.}
It satisfies
\eqn\wa{
\bar{S}^{n} S^{n}=I~,\quad S^{n} \bar{S}^{n} = I - P_n~,
}
where $I$ is the identity operator and
  $P_{n}$ denote the projection operators on to the first
$n$ states
\eqn\ea{P_n = \sum_{k=0}^{n-1} |k\rangle \langle k|~.}
Thus $U=S^n$ are solution generating transformations.

In the remainder of this  section we show how these abstract 
considerations work in some explicit examples. We will need
the following notation.  Coordinates commute according to 
\eqn\eb{[x^i,x^j]= i \Theta^{ij}.}
In $2+1$ dimensions we write $\th = \Theta^{12}$.  In higher dimensions
we skew-diagonalize $\Theta$ with $\th^i = \Theta^{i,i+1}$.  
Integrals over any two noncommutative
directions appear in the operator representation as
\eqn\wb{
{1\over 2\pi\theta}\int d^{2}x \to {\rm Tr}~.
}
In complex coordinates, $z = (x^1 + i x^2)/\sqrt{2}$, derivatives become
\eqn\a{\p = \p_z = -\th^{-\h}[\ad, \cdot]~, 
\quad \pb = \p_{\bar{z}}=\th^{-\h}[a,\cdot]~,}
where $a=z/\sqrt{\theta}$ so that 
\eqn\ab{[a,\ad]=I~.}
We write the gauge potential and field strength as
\eqn\d{\eqalign{A=&~A_z~, \quad \Ab=A_{\bar{z}}~, \cr
F =&~ iF_{z\bar{z}}=i\left(\p\Ab-\pb A-i[A,\Ab]\right) = 
\th^{-1}\left([C,\Cb]+I \right)~,}}
where
\eqn\CAdef{
C=\ad + i\th^{\h} A~, \quad  \Cb=a-i\th^\h \Ab~.
}
When considering noncommutativity in several complex dimensions 
simultaneously an additional index is introduced on the various symbols
to distinguish subspaces.

\subsec{Soliton Examples}
Consider a noncommutative gauge theory in $D=2+1$ dimensions 
\eqn\twodft{S = \int dt d^{2}x
\left( -{1 \over 4} (F_{\mu\nu})^2 + {1 \over 2}
D^{\mu} \phi D_{\mu} \phi - V(\phi-\phi_{\star}) \right)~.}
Our metric convention is $g_{\mu\nu}= (+,-,-)$. We take
the potential $V$ to have a local minimum at $\phi=\phi_*$
with $V(0)=0$ and a local maximum at $\phi=0$.
We take the scalar field in the adjoint representation so the
covariant derivative is
\eqn\fa{D_{\mu} \phi = \p_{\mu} \phi - i [A_\mu,\phi]~,}
or in the operator formalism
\eqn\fb{\eqalign{
D\phi \equiv &D_z \phi = -\th^{-\h}[C,\phi]~, \cr
  \Db \phi \equiv &  D_{\bar{z}}\phi =
\th^{-\h} [\Cb,\phi]~.
}}
The simplest solutions to the theory \twodft\ are spatially uniform, 
with the scalar field equal to an extremum of the potential, 
$\phi=\phi_{\star}$, 
and the gauge field vanishing, {\it i.e.} $C= \ad $ and ${\bar C}=a$. 
In \twodft\ the scalar field has been shifted such that the Lagrangian 
has no terms linear in the fields. This is important because otherwise 
the equations of motion would have a term proportional to the identity 
operator $I$ which transforms in the identity representation (for 
non-unitary transformations the adjoint representation would take 
$I\to UI{\bar U}=P$). Thus each term in the equations of motion 
transforms in the adjoint of $U({\cal H})$ so, without detailed 
computations it is clear that they are invariant under ``gauge'' 
transformations $\phi\to U\phi \bar U$, $C\to UC \bar U$
even for  non-unitary $U$ satisfying
\runi\ and \luni. Taking $U=S^{n}$  and using \wa\ we generate the 
soliton solutions
\eqn\twodsol{\eqalign{
\phi=&\phi_{\star}(I-P_{n})~,\cr
C=&S^{n} \ad \bar{S}^{n}~,\cr
{\bar C} =& S^{n}a \bar{S}^{n}~.
}}
Note that here and in all other solutions considered in this paper the 
barred gauge field ${\bar C}$ is simply the hermitian conjugate of $C$.

In position space the operator $P_n$ vanishes exponentially 
outside a region of linear extent $\sim\sqrt{n\theta}$; so the 
tachyon field in \twodsol\ is excited from the vacuum 
$\phi=\phi_\star I$ in a region of this size. The field strength 
\eqn\gauf{
F = {1\over\theta}\left([C,{\bar C}] + I\right) = 
{1\over\theta}~P_n~,
}
is similarly localized, so $C=\ad$ asymptotically.
 The solution \twodsol\ is thus interpreted
as a well-localized soliton.

In operator form the energy of static solutions to the theory \twodft\ 
is
\eqn\wc{
E = 2\pi\theta~{\rm Tr} \left({1\over 2}F^{2}+{1\over\theta}[C,\phi][{\bar 
C},\phi] + V(\phi-\phi_{\star})\right)~.
}
We normalize the potential so that the state far from the soliton
has vanishing energy 
$V(0)=0$ (otherwise the energy would diverge). Then the solution 
\twodsol\ has energy 
\eqn\wdd{
E = 2\pi\theta n\left( {1\over 2\theta^{2}} + V(-\phi_{\star})\right)~.
}

As far as we are aware the soliton solutions \twodsol\ are new.
At infinite noncommutativity the 
gauge field is negligible and our solutions reduce to the pure scalar field 
solutions of \gms.   
 For $\phi_\star =0$  they reduce to the pure 
gauge theory solutions found in \poly\ and studied in detail in
 \agms.  Note that the energy \wdd\ diverges as $\th \rightarrow 0$,
so we do not find well behaved solutions in the commutative theory.

A variation of  the construction above is to reconsider the
theory \twodft\ but taking
instead the scalar field $\phi$ in the fundamental representation of the 
gauge group. Then $\phi\to U\phi$ under gauge transformations. Following 
the steps above we find solutions with 
\eqn\we{
\phi = S^{n}\phi_{\star}~,
}
and the gauge fields given in \twodsol.  For a quartic potential, this 
solution was first found in \dbak\ and represents a vortex.  Generalizations
are considered in the following section.

\subsec{Vortex Examples}
Motivated by brane constructions it is natural to consider theories
with a complex scalar field $\phi$ transforming in the bi-fundamental
of two $U(1)$ gauge groups. The Lagrangian
\eqn\w{S = \int \left( -{1 \over 4} (F^+_{\mu\nu})^2 
-{1 \over 4} (F^-_{\mu\nu})^2
+ {1 \over 4}(D^\mu \phi D_\mu \phib + D^\mu \phib D_\mu \phi) 
+ V(\phi\phib-1) + W(\phib\phi -1)\right)~,}
with covariant derivatives
\eqn\s{\eqalign{D_{\mu} \phi&= \p_\mu \phi- i(A_\mu^+ \phi - \phi A_\mu^-)~, \cr
D_{\mu} \phib &= \p_\mu \phib +i(\phib A_\mu^+  -  A_\mu^- \phib)~,
}}
is invariant under the $U({\cal H}) \otimes U({\cal H})$ gauge symmetry
\eqn\ra{\eqalign{
\phi\rightarrow &V \phi \bar U ~,\cr
\quad C^- \rightarrow &U C^-  \bar U~,\cr
\quad C^+ \rightarrow &V C^+ \bar V~.
}}
The non-unitary isometries  $U=S^{n}$, 
$V=S^{m}$ generate the solutions
\eqn\vorantvor{\eqalign{
\phi =& S^{m} \bar{S}^{n}~,\cr
C^{-} =& S^{n}\ad \bar{S}^{n}~,\cr
C^{+} =& S^{m}\ad \bar{S}^{m}~.
}}
The corresponding field strengths
\eqn\ta{
F^{-}= {1\over \theta}P_{n}~,\quad
F^{+}= {1\over \theta}P_{m}~,
}
identify the solutions \vorantvor\ as coincident vortices, charged 
with respect to both $U(1)$ fields. The flux is  quantized,
\eqn\taa{ \int \! d^2x \, F^\pm = 2\pi \th~ {\rm Tr}( F^\pm) = 
~2\pi  ({\rm integer}).}
In string theory these solutions
will be interpreted as vortex/anti-vortex configurations corresponding
to coincident branes and anti-branes. 

It can be verified by explicit computation that \vorantvor\ satisfy the
equations of motion. A first step is the operator expressions
\eqn\t{\eqalign{D\phi &= \th^{-\h}(-C^+ \phi + \phi C^-)~, \cr
D\phib &= \th^{-\h}(-C^- \phib + \phib C^+)~, 
}}
giving
\eqn\cc{D\phi=D\phib=\Db\phi=\Db \phib=0~.}
These relations are not surprising because they hold trivially in 
vacuum, and they are preserved by the solution generating transformation. 

In the limit of large noncommutativity a class of these solutions was first
found in \refs{\hklm,\witncsft}.    
Our solutions are singular in the $\th \rightarrow 0$  limit.  This is to 
be contrasted with the approximate vortex solutions studied in \jmw.  
The latter reduce to the standard Nielsen-Olesen vortex in the commutative
limit. 

\subsec{The ABS Construction}
The vortex solution can be generalized to solitons localized
in more dimensions. The relevant construction is due to
Atiyah-Bott-Shapiro (ABS) \abs\ and is well known from the construction 
of BPS solitons in the $D{\bar D}$-system \absapp. 
Our discussion follows the noncommutative generalization of these
solutions in \harmor\ where more details 
can be found (see also \refs{\absapp,\witncsft}).

We want to construct a soliton of co-dimension $2p$. The starting point 
is a set of gamma matrices $\Gamma_j$, $j=1,2,\ldots 2p$
which map $S_+$ to $S_-$ where
$S_{\pm}$ are the two 
$2^{p-1}$ dimensional spinor representations  
of the $SO(2p)$ rotation group transverse to the soliton. 
General principles give \harmor\ 
\eqn\kernels{\eqalign{
{\rm dim} ~{\rm ker}~\Gamma\cdot x =& 1~, \cr
{\rm dim} ~{\rm ker}~{\bar\Gamma}\cdot x =& 0~,
}}
with suitable conventions for $\Gamma$-matrices. It follows that
\eqn\gg{\bar{T} = \bar{\Gamma}\cdot x {1 \over \sqrt{\Gamma\cdot x~ 
{\bar\Gamma}\cdot x}}~,}
is well defined and satisfies
\eqn\hh{T \bar{T} =I~, \quad~ \bar{T} T = I - P_{1}~,}
where $P_{1}$ is the projection operator onto the one-dimensional kernel
of $\Gamma \cdot x$. It is now clear that $U=\bar{T}$ is a solution
generating transformation. The resulting gauge fields are 
\eqn\ii{ C^+_{\alpha} = \ad_{\alpha}~, \quad~ C^-_{\alpha} = 
\bar{T}\ad_{\alpha} T~,}
where the index $\alpha=1,\ldots,p$ ennumerates the subspaces.

Let us be more explicit for the case of $p=2$.  We label the two 2D 
subspaces by coordinates $z$ and $w$, and work in terms of operators acting 
on the Hilbert space of two by two matrices with entries in
 ${\cal H} \otimes {\cal H}$. The gauge fields 
\ii\ then become
\eqn\jj{C_z^+ = \ad \otimes I~, \quad C_w^+ = I  \otimes 
\ad~, \quad
C_z^- = \bar{T} (\ad \otimes  I) T~, \quad
C_w^-=\bar{T} (I  \otimes \ad) T~.}
Choosing the noncommutativity $\th$ in the two subspaces to be equal,
the corresponding nonvanishing field strengths are
\eqn\kk{F^-_{z\bar{z}} =F^-_{w\bar{w}}= -i\theta^{-1}P_{1}~.}
This is a self-dual field strength (see below). 

The construction of multi-vortex/anti-vortex solutions from
the previous section can be repeated in $2p$ dimensions. 
The operator $T$ has a one-dimensional 
kernel and the first equation in \hh\ implies that it is surjective,
${\rm im} T = S_-\otimes {\cal H}^{\otimes p}$. 
It follows that $T$ maps 
a one-dimensional subspace into its own kernel, a process that can 
be repeated with the result (this also follows from an index theorem 
\countb\ )
\eqn\ya{
{\rm dim}~{\rm ker}~ T^{n} = n~,
}
and so
\eqn\ll{T^{n}\bar{T}^{n} = I~,\quad \bar{T}^n T^n = I- P_n~,}
where $P_n$ is the projection operator onto the kernel of $T^n$.
We can thus use the operator $U=\bar{T}^{n}$ as a solution generator; moreover
there are two gauge fields transforming independently, as in \ra, 
so we can also use $V={\bar T}^m$. The result is
multi-ABS/anti-ABS solitons with the gauge fields 
\eqn\ij{ C^+_{\alpha} = \bar{T}^{m}\ad_{\alpha} T^{m}~, 
\quad~ C^-_{\alpha} = \bar{T}^{n}\ad_{\alpha} T^{n}~.}

We have several comments:
\item{(1)}
The manipulations above are quite general; they can be repeated 
after replacing $\Gamma\cdot x$ with any other operator satisfying 
\kernels. The operators $T$ and $\bar{T}$ are known as Toeplitz operators. 
\item{(2)}
The derivation illustrates a general feature of the solution 
generating technique, namely that we do not need an explicit action to 
find a solution, only some general properties of the theory.

\item{(3)}
The gauge field $C^{-}$ in \ii\ falls off like $1/r$ in position space.
This na\"{\i}vely indicates a field strength $F^-\sim 1/r^{2}$ and 
thus an infrared divergence in the energy after the
spatial integration for all $p>1$. However, the actual field 
strength, $\theta F^- = P_1$ ,
gives finite energy --- it corresponds in position space to a Gaussian
falloff. The derivative and commutator terms in the field 
strength apparently cancel all the power law dependence.
In generic commutative theories the $1/r$ falloff of the gauge field
(which is required for an acceptable falloff of the scalar field
derivatives) does lead to a divergent energy, and so eliminates the
possibility of solitons with more than three transverse directions\foot{See
e.g. \coleman. We 
thank G. Moore for bringing this to our attention.}.  So this gives another
instance of noncommutative solitons with no commutative counterparts.

\subsec{More on Instantons}
At this point it is natural to ask what class of noncommutative
solitons can be generated using our technique. 
To explain a
specific limitation we consider the important example of instantons
in a noncommutative $U(1)$ gauge theory in four dimensions \neksch. 
The equations
of motion
\eqn\csym{
[C^\mu,[C_\mu,C_\nu]]=0~,
}
where $\mu,\nu=1,2,3,4$ are invariant under $C_{\mu}\to UC_{\mu}\bar{U}$ 
for all the $U$ satisfying $\runi$, including those of the form $\luni$. 

Any self-dual or anti-self-dual field strength automatically
satisfies \csym.  In the operator formalism the self-duality condition is
\eqn\sdcond{\eqalign{
[C_{z},{\bar C}_{\bar w}]=&0~,\cr
[C_{z},{\bar C}_{\bar z}]=&[C_{w},{\bar C}_{\bar w}]~,
}}
and the anti-self-duality condition reads
\eqn\asdcond{\eqalign{
[C_{z},C_{w}]=& 0~,\cr
[C_{z},{\bar C}_{\bar z}]+[C_{w},{\bar C}_{\bar w}]=&-2I~.
}}
We chose complex coordinates in two 2D subspaces with identical
noncommutativity, so $\Theta$ is self-dual; 
furthermore $\epsilon_{zw{\bar z}{\bar w}}=1$.
Now, it is simple to show that the self-duality condition \sdcond\ is 
invariant under $U$ transformations satisfying \runi, including those satisfying 
\luni, whereas the anti-self-duality 
condition \asdcond\ is invariant only if also $\bar U U=I$, {\it i.e.} 
under true gauge transformations. Self-dual instantons are thus simple to 
generate because the self-duality conditions are 
preserved under solution generating transformations. Indeed, 
starting with vacuum and transforming with a suitable $U$ we return to
\kk, recovering the self-dual instantons of \agms. On the other hand, it 
is more difficult to generate the anti-self-dual instantons of Nekrasov and 
Schwarz \neksch\ since 
transformations of the vacuum are generally not anti-self-dual. Of course 
this can also be turned into an advantage: starting with the 
Nekrasov-Schwarz anti-self-dual instanton we can generate many other 
solutions that are similarly unconnected to the vacuum.

\newsec{Applications in String Field Theory}

So far we have discussed noncommutative Yang-Mills-Higgs theories, but
it should be clear that our method applies much more generally.  The key
feature is the existence of $U({\cal H})$ gauge symmetry.  In this section
we study actions arising from string field theory, and our exact solutions 
will correspond to D-branes.  As discussed in section 4, this 
generalizes the construction of \hklm, and shows that the 
results obtained there are exact for any value of $\th$.  

The first issue we need to discuss is the realization of gauge symmetry
in string field theory.  Witten's cubic open string field theory 
\witcub\
is invariant with respect to the gauge transformations 
\eqn\za{\delta_\Lambda \Psi = Q\Lambda + g_s\Psi *\Lambda - 
g_s \Lambda * \Psi~,}
where $\Lambda$ is a ghost number zero
string field.  The $g_s$ dependent contribution
mixes component fields of different levels.  The resulting transformations
look very unusual and unwieldy
 when expressed in component form, and the action truncated
to any finite number of components is not gauge invariant.   For example,
the lowest level component of $\Lambda$ induces the following (schematic)
 transformations on the tachyon and gauge field \ks:
\eqn\zb{\eqalign{\delta_\lambda \phi =~& [\lt,\Tt] + \{\lt,\at\} +
\{\partial_\mu \lt
,\At^\mu\} +\{\lt,\partial_\mu \At^\mu\}~ + ~\cdots~,  \cr
\delta_\lambda A_\mu =~& \partial_\mu \lambda + [\lt,\At_\mu]
+ [\at,\partial_\mu \lt] +[\lt,\partial_\mu \at] + \{\lt,\partial_\mu \Tt \}
 + \{\partial_\mu \lt,\Tt\} \cr 
&+ [\partial_\mu \lt,\partial^\nu \At_\nu]+
[\lt,\partial_\mu \partial^\nu \At_\nu] 
+ [\partial^\nu\partial_\mu \lt,\At]
+[\partial^\nu \lt, \partial_\mu \At_\nu]~ + ~\cdots~.}}
Here $a$ 
is an auxilliary scalar, $\cdots$ indicate contributions of higher
level fields, and $\tilde{f} = \exp{\left[\alpha' \ln(3\sqrt{3}/4)
\partial_\mu \partial^\mu \right]} f~$ for any function $f$.  Also, 
Chan-Paton factors have been included and so all fields are $N \times N$ 
matrices.  

To obtain a simpler realization of the gauge invariance we can imagine
integrating out classically all fields except for the tachyon and gauge
field.  Then it is believed \david\
 that there exists a field redefinition such that 
\eqn\zc{\eqalign{\delta_\lambda \phi =~& -i[\lambda,\phi]~,\cr
\delta_\lambda A_\mu =~& \partial_\mu \lambda - i[\lambda,A_\mu]~.}}
In particular, in the presence of noncommutativity there will be a 
$U({\cal H})$ gauge symmetry with fields transforming as in the examples
of the previous section.  It is then simple to apply our solution 
generating technique to arbitrary gauge invariant actions expressed in 
terms of these variables.  This is the approach we follow in this
section.

As is well known from the work of \refs{\schom,\sw}, the B-field is
incorporated into the action by replacing ordinary products by 
$\star$ products, and the closed string metric $g_{\mu\nu}$ and
coupling $g_s$ by the open string metric $G_{\mu\nu}$ and coupling
$G_s$.  There is some freedom in the choice of the open string
quantities corresponding to a choice of $\Phi$ parameter; see
\nati\ for detailed discussion.  For our purposes, it is 
convenient to take $\Phi = -B$.  In the case of maximal rank B-field
and Euclidean signature this implies the relations  
\eqn\zc{\eqalign{\Theta=& {1 \over B}~, \cr
G =& -(2\pi \alpha')^2 B{1 \over g}B~, \cr
 G_s =&~ g_s \det(2\pi\alpha' Bg^{-1})^\h~.}}
The  noncommutative field strength $F$ only appears in the action 
in the combination 
$F+\Phi$, or equivalently given our choice for $\Phi$, only through $[C,\Cb]$.

On the other hand, for constructing codimension $2p$ solutions it is
only necessary to turn on noncommutativity in $2p$ directions.  For
simplicity we will explicitly consider $p=1$, but the generalization to
arbitrary $p$ is straightforward.  In $d+1$ Minkowskian  dimensions we take
$g_{\mu\nu}=\eta_{\mu\nu}$ and
\eqn\zfa{B_{d-1,d} = b, \quad b<0~,}
so that
\eqn\zfb{\eqalign{\theta \equiv&~ \Theta^{d-1,d} = {1 \over |b|}~, \cr
  G_{\mu\nu}=&~ {\rm diag}(1,-1,\ldots,-1,-(2\pi\alpha' b)^2,
-(2\pi\alpha' b)^2)~, \cr
G_s =&~ (2 \pi \alpha' |b|)g_s~.}}
It is convenient to use complex coordinates in the noncommutative 
directions,
\eqn\zcd{z = {1 \over \sqrt{2}}(x^{d-1} + i x^{d})~.}
In the operator representation we introduce $C$ and $\overline{C}$ as in
section 2.

\subsec{Bosonic string}
The action for the tachyon and gauge field in the absence of the B-field is
\eqn\zda{S = {c \over g_s} \int\! d^{26}x \sqrt{g}\left\{
-{1 \over 4}h(\phi-1)F^{\mu\nu}F_{\mu\nu}   + \cdots +     
{1 \over 2}f(\phi-1)\partial^\mu \phi \partial_\mu \phi + \cdots - V(\phi-1)\right\}.}
Omitted terms indicate higher powers of fields and higher derivatives.
 It is convenient for us to take the arguments of $h,f$ and $V$ to be $\phi-1$.
 We have defined $\phi$
so that $V(\phi-1)$ has a local maximum at $\phi=0$ with $V(-1)=1$ 
(the $D25$-brane),
 and a local minimum at $\phi=1$ with $V(0)=0$ (the closed string vacuum).  
With these conventions, the $D25$-brane tension in the absence of B-field
 is $T_{25}= c/g_s$.   
The functions $h(\phi-1)$ multiplying powers of $F_{\mu\nu}$ without derivatives
are all known
since they sum up to give $V(\phi -1)$ times the Born-Infeld action. 
In particular, these functions all vanish in the closed string vacuum
 \refs{\senvanish,\hkm,\bsft}; 
as will see, this will be important when we come to
 verifying that our solitons have the correct tensions to be identified
with D-branes.  

Now we turn on the B-field in $x^{24}$, $x^{25}$.  The tachyon 
transforms in the adjoint of noncommutative U(1) as in \fa, and the 
field strength becomes
\eqn\zfe{F_{\mu\nu} = \partial_\mu A_\nu - \partial_\nu A_\mu 
+i[A_\mu,A_\nu]~.}
Using the operator representation in the noncommutative directions, the
action becomes,
\eqn\zff{\eqalign{S = {2 \pi \th c \over  G_s} \int\! d^{24}x 
\sqrt{G}~{\rm Tr}
&
\left\{ 
-{1 \over 4}h(\phi-1)(F^{\mu\nu}+\Phi^{\mu\nu})(F_{\mu\nu}+\Phi_{\mu\nu})
   + \cdots 
\right. 
\cr
 &~~~+ 
\left.    
{1 \over 2}f(\phi-1)D^\mu \phi D_\mu \phi + \cdots - V(\phi-1) 
\right\}
~,}}
with
\eqn\zfg{F_{24,25}+\Phi_{24,25} = -iF_{z\overline{z}} + {1 \over \th}
= -{1 \over \th}[C,\Cb]~.}
   All indices are contracted with $G_{\mu\nu}$, and $\star$ products are
implied.  Also, we have not specified the factor ordering
in \zff\ since it will not be necessary to do so. 

Now we repeat the argument from section 2.  Let $U \in U({\cal H})$ satisfy
$\bar U U = I$ and be independent of $x^0, \cdots, x^{23}$. 
 The equations of motion 
following from \zff\ are invariant under the transformations 
\eqn\ze{\eqalign{\phi \rightarrow & ~U \phi \bar U~, \cr
C \rightarrow & ~U C \bar U~,\cr
A_\mu \rightarrow & ~U A_\mu \bar U~, \quad \mu =0 \ldots 23~. }}
Taking $U=S^n$ so that
\eqn\zg{ U \bar U = I - P_n~,}
and acting on the closed string vacuum --- 
$\phi=1$, $C=\ad$, $A_\mu=0$ --- we thus generate the  solutions 
\eqn\zf{\eqalign{\phi =&~ S^n \bar{S}^n = (I-P_n)~, \cr
C =&~ S^n \ad \bar{S}^n~, \cr
A_\mu =&~0 , \quad \mu =0 \ldots 23~.
}}
They are interpreted as $n$ coincident $D23$-branes.

It is easy to evaluate the energy of this solution for the general action
\zff\ because the only contribution comes from the potential term.
In particular, all terms involving covariant derivatives of $\phi$ or
$F_{\mu\nu}$ vanish because they do so in the closed string vacuum and
this is preserved by the solution generating transformation.  We also
have 
\eqn\zfg{h(\phi-1)[C,\Cb]^2 = h(-P_n)(1-P_n) = h(-1)P_n (1-P_n)=0~.}
In the second step we used $h(0)=0$.   Terms with higher powers of
$[C,\Cb]$ similarly vanish (with any choice of factor ordering) simply 
because the projection operators $P_n$ and $1-P_n$ are orthogonal. For
the potential term we have
\eqn\zfh{V(\phi-1) = V(-P_n) = V(-1)P_n = P_n~.}
Then using 
\eqn\pp{ {\sqrt{G} \theta \over G_s} = {2\pi \alpha' \over g_s}~,}
we find for the action \zff\ evaluated on our solution,
\eqn\zfi{S = {(2\pi)^2 \alpha' n c \over g_s}\int\! d^{24}x~,}
which identifies the tension as
\eqn\zfj{ T =  {(2\pi)^2 \alpha' n c \over g_s} = (2\pi)^2\alpha' n T_{25}
 = n T_{23}~,}
as desired.  

Repeating this construction for $2p$ noncommutative directions we find
$D(25-2p)$-branes with the correct tension.  

The construction we have given here reduces to that of \hklm\ in the
limit of large B-field.  There, the effect of the large B-field was to
suppress the derivative contributions.  Here, we have an exact construction
for any value of B.  The key point is that the solution generating
transformation generates a gauge field which sets the covariant 
derivatives of fields to zero.

\subsec{The Superstring}

In type II string theory there are two types of unstable Dp-branes:
``wrong'' $p$ non-BPS $Dp$-branes, and the  $D\overline{D}$ system.  
The construction of solitons on the former class of D-branes is a trivial
extension of our discussion for the bosonic string, so we focus here
on the $D\overline{D}$ system.   The field content of the effective
theory after integrating out massive modes corresponds to the theory
studied in section 2.3.  
The tachyon potential is 
a ``Mexican hat''  with minima at $|\phi|=1$. There is also a symmetry
given by the action of $(-)^{F_L}$ which corresponds to interchanging
the $D$ and $\overline{D}$ \sennon:
\eqn\zga{ (-1)^{F_L}:\quad
   \phi \leftrightarrow \phib, \quad A^+ \leftrightarrow A^-~.}

Next we write down an action for the string field theory, generalizing
\w\ and satisfying some basic properties.  First consider the gauge
kinetic terms.  Given the symmetry \zga\ we write
\eqn\zgb{{\cal L}_{gauge} = h_+(|\phi|^2-1)(F^-_{\mu\nu}+F^+_{\mu\nu})^2
+ h_-(|\phi|^2-1)(F^-_{\mu\nu}-F^+_{\mu\nu})^2.}
The function $h_+$ is known to vanish in the closed string
vacuum.

In the noncommutative
theory $F^-$ and $F^+$ transform differently under the two $U(1)$ 
factors, but one can form linear combinations by noting that 
$F^-$ and $\phib F^+ \phi$ transform in the same way.  So an acceptable 
gauge kinetic term is 
\eqn\zgc{\eqalign{{\cal L}_{gauge} 
=&~ h_+(\phib \phi-1)\left\{F^-_{\mu\nu}+ \Phi_{\mu\nu}
+\phib( F^+_{\mu\nu}+ \Phi_{\mu\nu})\phi \right\}^2  \cr
+&~ h_-(\phib \phi-1)\left\{F^-_{\mu\nu}+ \Phi_{\mu\nu}
-\phib( F^+_{\mu\nu}+ \Phi_{\mu\nu})\phi \right\}^2 \cr
 +&~
 \left\{\phi \leftrightarrow \phib, \quad A^+ \leftrightarrow A^-\right\},}}
with $h_+(0)=0$. 
The last line is included for symmetry under $(-)^{F_L}$.  A similar
expression holds for terms including higher powers of the gauge fields.

Tachyon kinetic terms appear as in \w, but now multiplied by functions
$f(\phib \phi-1)$ and symmetrized.  Symmetry under $(-)^{F_L}$ implies that the potential is
of the form
\eqn\zgd{V(\phi\phib-1)+V(\phib \phi-1)~.}
In the notation of \w\ we take $V=W$.

Now we use our solution generating transformation to construct exact
solitons representing BPS D-branes.  As before, we will explicitly
consider the codimension two case; starting with a spacefilling
$D9-\overline{D9}$ system of IIB this will produce  BPS $D7$-branes.
As in \vorantvor, the solution we generate starting from the closed string
vacuum is
\eqn\vorantvorb{\eqalign{
\phi =&~ S^{m} \bar{S}^{n}~,\cr
C^{-} =&~ S^{n}\ad \bar{S}^{n}~,\cr
C^{+} =&~ S^{m}\ad \bar{S}^{m}~, \cr
A^+_\mu =&~ A^-_\mu =0~, \quad \mu =0 \ldots 7~.
}}
We claim that this solution represents $m$ $D7$-branes coincident
with $n$ $\overline{D7}$-branes. 

We now work out the energy of this solution.  As in the bosonic case,
covariant derivatives of the tachyon and field strengths 
vanish before and after the
transformation, and so do not contribute to the energy.  It is less
trivial to verify that the gauge field terms \zgc\ do not contribute.
We need to compute
\eqn\zgd{h_+(\phib \phi -1)
\left\{[C^-,\overline{C^-}] + \phib[C^+,\overline{C^+}]\phi
\right\}^2 +
h_-(\phib \phi-1)\left\{[C^-,\overline{C^-}] - \phib [C^+,\overline{C^+}]\phi
\right\}^2.}
For our solution,
\eqn\zge{\eqalign{\phi \phib =&~ I-P_m~, \cr
\phib \phi =&~  I-P_n~, \cr
[C^-, \overline{C^-}]=&~-(I-P_n)~,\cr
 \phib[C^+, \overline{C^+}]\phi=&~-(I-P_n)~.}}
The first term in \zgd\ vanishes since $h_+(\phib \phi-1)=h_+(-P_n)
=h_+(-1)P_n$, which is orthogonal to $I-P_n$;  here we used that
$h_+(0)=0$.   The second term in \zgd\ vanishes without a similar assumption
about $h_-$.   So as in the bosonic theory, the only contribution to
the energy comes from the potential term, which we find to be 
\eqn\zgf{V(\phi\phib-1)+V(\phib \phi-1) = V(-P_m)+V(-P_n) = V(-1)(P_m +P_n).}
Repeating the computation leading to \zfj\ in the bosonic case now
gives the tension
\eqn\zgg{T_{nm} = {(2\pi)^2 \alpha' (n+m) c \over g_s}T_9 = (n+m)T_7.}
as expected for $m$ $D7$-branes plus $n$ $\overline{D7}$-branes.

Using the ABS construction of solitons in section 2.4, it is 
straightforward to generalize the above discussion to codimension
$2p$ solitons representing coincident $D(9-2p)$ and 
$\overline{D(9-2p)}$ branes.

\newsec{Discussion}

In this paper we have constructed soliton solutions of noncommutative
gauge theories with or without scalar (tachyon) fields for arbitrary
values of the noncommutativity parameter $\theta$. In the context
of an effective low-energy description of string field theory these
solutions represent various D-brane solutions discussed 
previously at infinite noncommutativity in 
\refs{\dmr,\hklm,\witncsft}. 
 We would now like
to discuss some of the implications of these solutions and the technique
used to generate them.

One issue which has arisen in discussions of noncommutative tachyon
condensation involves the correct description of the closed string
vacuum. Clearly in the vacuum the tachyon field $\phi$ takes on its minimum
everywhere, $\phi= \phi_* I $, but various possibilities have been discussed
for the gauge field configuration. Given the vanishing of the gauge kinetic
term in the closed string vacuum, these different choices are degenerate in
energy at the level of effective field theory. It was argued in \senvac,
based in part on studies in truncated string field theory \somerefs,
that these choices are in fact illusory, an artifact of using the wrong 
variables to describe the closed string vacuum.

In \refs{\hklm,\witncsft} one obtains a simple description
of $D$-branes as noncommutative solitons by taking a limit of
large noncommutativity, the key point being that derivative corrections
are suppressed in this limit.  In \hklm\ it was suggested that this
description should also hold at finite or vanishing noncommutativity
since a purely transverse B-field can  be shifted by a gauge 
transformation, but it was not clear how to show this explicitly
given the presence of derivative corrections.    Since the limit also
requires one to take the closed
string coupling $g_s \rightarrow 0$ in order to obtain a finite
open string coupling, another concern was whether this limit of the
theory is smoothly connected to the commutative theory with finite
but small $g_s$.\foot{Similar issues arise in the proposal of
\wittenstrings.} Various discussion of these issues can be found in
\refs{\gmsii,\nati,\senvac,\krs}.

To resolve these issues one should distinguish several
possible notions of ``$\theta$'' independence. In string theory
the gauge invariant physical quantity 
is\foot{Here $F$ is the commutative field strength. The 
noncommutative field strength denoted $F$ in the main text is changed 
to ${\hat F}$ in the discussion section, conforming with the notation 
of \sw.} 
${\cal F} = F+B$.
 Different choices of $F$ and $B$ with fixed ${\cal F}$ lead to 
different but gauge equivalent descriptions of the physics. The physical issue
is one of the ${\cal F}$ dependence of the D-brane tensions and spectra
computed using noncommutative solitons. 

The  noncommutative description of D-branes involves the noncommutative
field strength $\hat F$ and a two-form $\Phi$ \refs{\pioline,\sw,\nati}.
In this language the issue is one of dependence on the physical
combination $\hat F + \Phi$. The discussion of \nati\ makes it clear
that one can formulate the theory in a form which is independent of
the background value of this physical combination. In this language
one must choose a classical solution for the quantities
$X^i = \Theta^{ij}C_j$ with specified boundary conditions at infinity.
Specifying boundary conditions is equivalent to choosing a vacuum,
$X^i=x^i$ with $[x^i,x^j]=i \Theta^{ij}$, and the question of 
$\theta$ independence is whether the physics is independent of this choice. 
Both \hklm\ and \gmsii\ were based on the vacuum $X^i=0$: 
  \hklm\ phrased
this in terms of the limit $B \rightarrow \infty$ with $\Theta = 1/B$,
while \gmsii\ set $C_i =0$.\foot{More precisely, \hklm\ took 
$B \rightarrow \infty$ in transverse directions,
while \gmsii\ took $C_i=0$ in all directions including time.}
Here we have shown that the correct properties
of D-branes are recovered for any choice of vacuum, that is
\eqn\convac{
X^i = \left({1\over B}\right)^{ij}C_j~,\quad {\rm with}~
{1 \over \sqrt{2}}(C_{2\alpha-1}-iC_{2\alpha}) = \bar{a}_\alpha~,
}
for any $B$. 
Since D-brane properties are here found to be $\th$ 
independent, this result also supports the argument of \senvac\ that 
the superficially different vacua labelled by $\theta$ should actually
be identified.

Since we can in particular take $\theta$ finite, 
we are not forced to take $g_s \rightarrow 0$ in order to make the 
open string coupling finite. Thus the representation of D-branes 
as noncommutative solitons 
we have found is valid for arbitrary $\theta$ and small but non-zero 
$g_s$. This establishes that the solutions found in 
\refs{\hklm,\witncsft} are not just a description of D-branes 
in some unphysical regime of infinite noncommutativity and zero 
coupling, but rather are describing D-branes in the conventional 
weakly coupled string theory vacuum as argued in \hklm.  

Another issue raised by this work is related to the presence of non-zero
gauge fields in the exact solution, even in the bosonic string.
This would na\"{\i}vely seem to be in contradiction with the description of
D-branes as lumps in open bosonic string field theory \refs{\tsolrefs}
where  one argues that the gauge field can consistently be
set to zero in level truncation \refs{\ks,\truncate}. 
On the other hand, the
relation between the gauge invariant description used here and the
gauge fields which appear in string field theory is highly
non-trivial \david. It may be that a field redefinition and/or
string field theory  gauge transformation relates the two na\"{\i}vely
different descriptions. In type II string theory the question can be
asked more sharply since the gauge field strength acts as a source
of RR charge and hence must be non-zero in any construction of BPS
D-branes. To our knowledge the form of the gauge field for BPS
D-branes has not yet been examined in truncated open string field
theory. 

Finally, it would be interesting to apply these techniques directly
in string field theory. Cubic open string field theory \witcub\
even at $B=0$ has an infinite dimensional symmetry group, and it may
be possible to construct the analog of non-unitary isometries. It has
been suggested \schnabl\ that one might construct the closed string
vacuum in the form
\eqn\sch{A_0 = \bar V * Q V~,}
for some ``non-unitary isometry'' obeying $\bar V * V = I.$ \foot{
This proposal was also made by G. Moore
at Aspen in August, 2000.} Our considerations suggest that whether or
not this is the case, one should try to construct D-brane solutions
in string field theory by acting with non-unitary isometries on the
closed string vacuum state
\eqn\strd{(Q+A_{\rm D-brane}) =  V * (Q+A_0) \bar{V}~.}
It would be interesting to try to carry out this proposal concretely,
either in the theory with $B=0$ or in the theory with $B \ne 0$ where
the star product is modified to include the Moyal product on the
zero mode wave functions.

\bigskip\medskip\noindent 
{\bf Acknowledgements:}
We thank D. Kutasov, E. Martinec and G. Moore for discussions.
This work was supported in part by NSF grant PHY-9901194 and by DOE grant
DE-FG0290ER-40560. FL was supported
in part by a Robert R. McCormick fellowship. 

\listrefs

\end